\documentclass[letterpaper,twocolumn,fleqn]{article} 

\usepackage{ist}
\usepackage{biblatex} 
\usepackage{cuted}
\usepackage{capt-of}
\usepackage{pifont}
\usepackage{graphics}
\usepackage{amsmath}

\title{Hair Color Digitization through Imaging and Deep Inverse Graphics}

\author{Robin Kips$^{1, 2}$,
Panagiotis-Alexandros Bokaris$^1$,
Matthieu Perrot$^1$,
Pietro Gori$^2$,
Isabelle Bloch$^3$\\
$^1$ L'Or\'eal Research and Innovation, France\\
$^2$ LTCI, T\'el\'ecom Paris, Institut Polytechnique de Paris, France\\
$^3$ Sorbonne Universit\'e, CNRS, LIP6, Paris, France\\
}

\date{} 

\begin{document} 

\maketitle 

\thispagestyle{empty} 


\begin{strip}\centering
\includegraphics[width=1.0\linewidth]{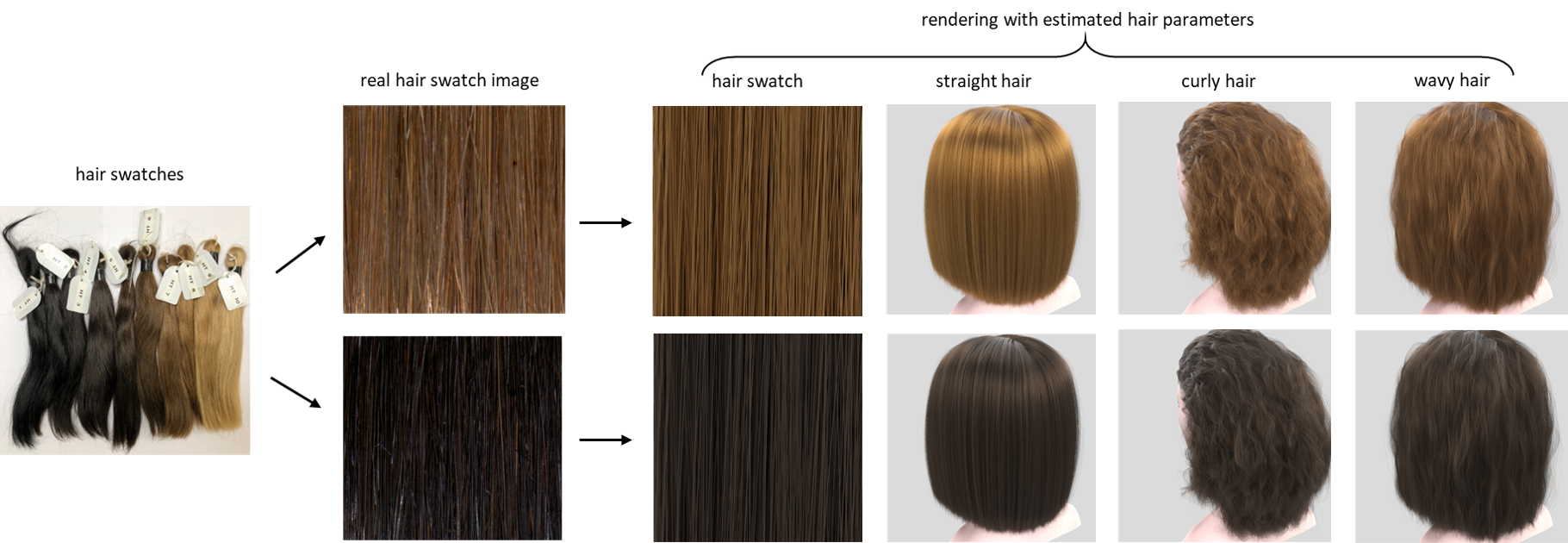}
\captionof{figure}{Given a hair sample, we propose a hair digitization method that estimates rendering parameters that can be used to render a synthetic scene with hair of similar color appearance. Our method is based on the combination of a controlled imaging method, a deep inverse graphics encoder model, and a path-tracing renderer. (Hair models courtesy of Cem Yuksel~\cite{Yuksel})}
\label{fig:header_figure}
\end{strip}

\begin{abstract}
Hair appearance is a complex phenomenon due to hair geometry and how the light bounces on different hair fibers.
For this reason, reproducing a specific hair color in a rendering environment is a challenging task that requires manual work and expert knowledge in computer graphics to tune the result visually. 
While current hair capture methods focus on hair shape estimation
many applications could benefit from an automated method for capturing the appearance of a physical hair sample, from augmented/virtual reality to hair dying development.
Building on recent advances in inverse graphics and material capture using deep neural networks, we introduce a novel method for hair color digitization. 
Our proposed pipeline allows capturing the color appearance of a physical hair sample and renders synthetic images of hair with a similar appearance, simulating different hair styles and/or lighting environments.
Since rendering realistic hair images requires path-tracing rendering, the conventional inverse graphics approach based on differentiable rendering is untractable. 
Our method is based on the combination of a controlled imaging device, a path-tracing renderer, and an inverse graphics model based on self-supervised machine learning, which does not require to use differentiable rendering to be trained. 
We illustrate the performance of our hair digitization method on both real and synthetic images and show that our approach can accurately capture and render hair color.
\end{abstract}




\section{Introduction}
Hair realism is required in numerous applications including but not limited to animation, gaming, special effects and virtual/augmented reality. Hair appearance complexity lies in the geometry and properties of hair fibers and the interaction between them. For this reason, defining and controlling hair appearance is difficult to automatize.

\paragraph{Hair rendering}
Various scattering models have been proposed in the literature to simulate the way light is reflected from hair fibers~\cite{dEon11, Marschner03, Yan15} offering physically-based realistic results. Different light bounces on the hair fiber, which is usually represented by a cylinder, have an important effect on the hair appearance and thus its realism. Specularities, hair color and hair tone are directly connected with the three scattering components introduced in~\cite{Marschner03}.
With the recent advances in the parallelization of ray tracing (\textit{OptiX} \cite{Optix}, \textit{Vulkan RT} \cite{Vulkan}), path-tracing implementations of these models have become more practical, increasing the degree of realism due to global illumination. 
The control over the appearance of the hair color is based on physical parameters of individual hair fibers, such as natural melanin concentration/ratio or other artificial dye colorants. In order to improve the user friendliness in production, visual attributes such as the albedo have been introduced in~\cite{Chiang15} and implemented in \textit{Renderman}~\cite{Renderman} to favor the artistic expression. However, the tuning of these parameters to match the appearance of a physical hair sample is almost impossible without a psycho-visual match.

\paragraph{Hair Capture}

Various methods have been proposed to capture hair appearance from example images. The general objective consists in estimating hair descriptors that can render hair with a similar appearance in a synthetic environment. 
The hair capture is often performed using a multiview system as in~\cite{luo2012multi} or RGB-D camera as shown in~\cite{luo2012multi}.
Other methods focus on conventional single view images to provide a more scalable solution to this problem, using databases of references 3D hair models for comparison as done in \cite{hu2014robust, Hu2015}.
However, all these methods focus on estimating hair shape and do not provide solutions to reproduce the color appearance of a hair, which is a challenging problem. Thus, most systems require hair color parameters to be set manually by expert artists. 
Furthermore, existing methods can only capture a complete hair style appearance and cannot be used on hair samples such as a single strand. 
In practice, many applications such as hair dye development would benefit from a hair capture system on a hair sample. This would allow to dye hair strands samples, avoiding testing dye formulas on the entire hair of a volunteer. 
To the best of our knowledge, there is no existing approach for capturing hair color appearance from hair strand images. 

\paragraph{Inverse graphics}
Given a natural image, \textit{inverse graphics} approaches aim to estimate features that are typically used in computer graphics scene representation, such as HDR environment map~\cite{somanath2020hdr} or meshes of 3D objects such as faces~\cite{FLAME:SiggraphAsia2017}. 
This idea has been successfully applied to material capture tasks, 
accelerating the computer graphics creation pipeline by estimating material parameters from an example image.
Inverse graphics models are generally based on a neural network that is trained on synthetic images using a differentiable renderer for supervision~\cite{Che2020}.
However, hair rendering requires path-tracing operations that are not supported by current differentiable renderers. Ray-tracing rendering is \textit{a priori} a non-differentiable operation. The gradients can be estimated approximately as in~\cite{Li18} but due to the amount of light bounces in hair geometry the implementation of a conventional inverse graphics approach is untractable in our case. 
Recently, a self-supervised approach for training an inverse graphics model was introduced in \cite{Kips_2021_CVPR}, relaxing the need for a differentiable renderer. 

In this paper, we propose to
build on recent advances in inverse graphics and material capture using deep neural networks, introducing a novel method for hair color digitization.
Given a physical hair sample, our objective is to build an automated method to capture the hair color appearance and render it in a synthetic environment automatically. Such a method has direct applications in virtual/augmented reality, as well as hair dying development.
Furthermore, we focus on hair color and exclude the estimation of hair shape parameters, which can be controlled \textit{a posteriori} in the rendering environment to simulate various hairstyles. 
In addition, color heterogeneity of different fibers or the amount of specular reflection (hair damage, sebum levels) has a direct effect on hair appearance, but capturing these effects are out of the scope of this paper.

\begin{figure}[!t]
\begin{center}
   \includegraphics[width=0.4\linewidth]{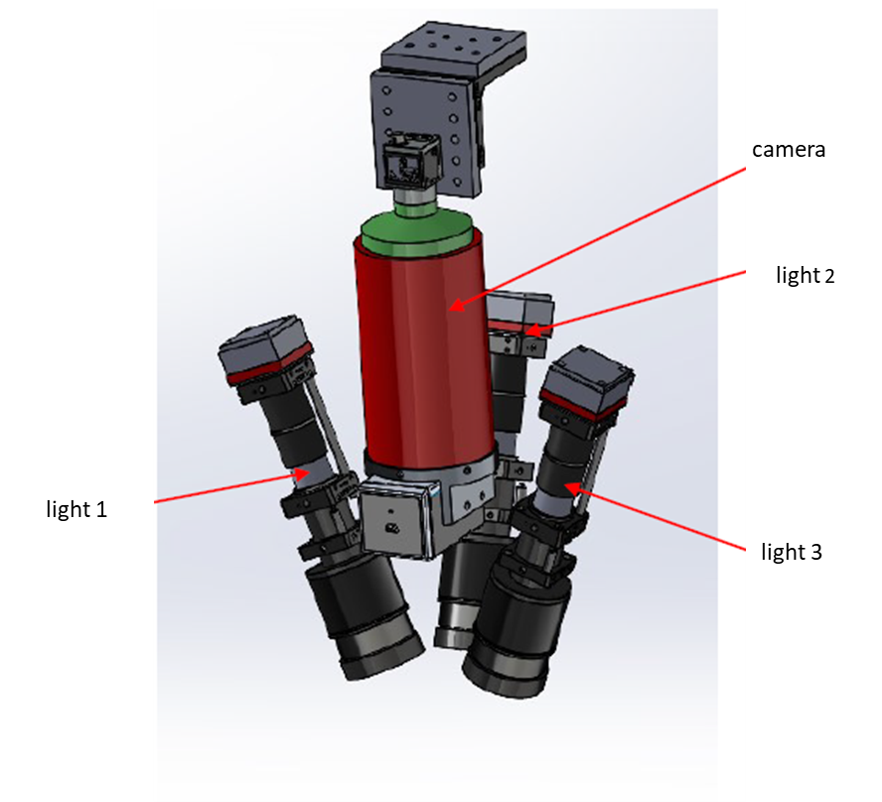}
\end{center}
   \caption{The imaging setup used for capturing hair sample images. It is composed of one camera and three light sources for controlled acquisitions conditions.}
\label{fig:image_device}
\end{figure}

\begin{figure}[!t]
\begin{center}
   \includegraphics[width=0.8\linewidth]{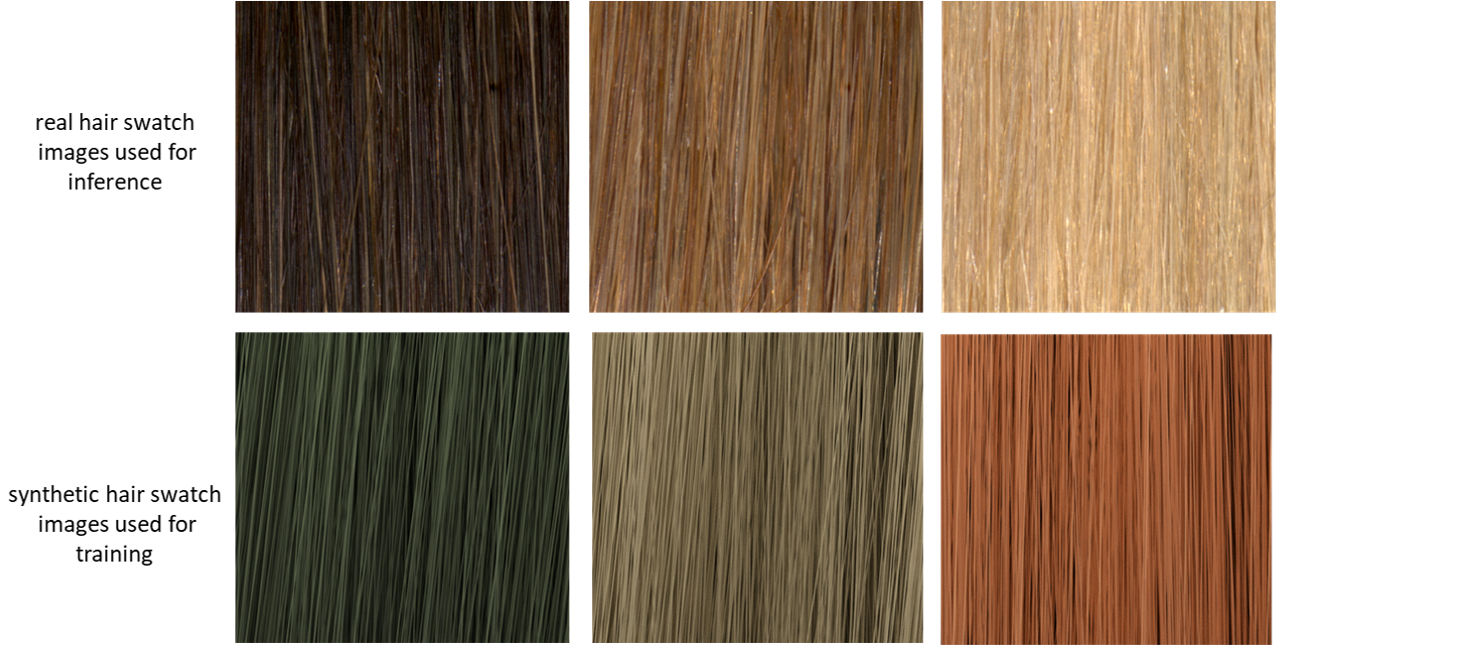}
\end{center}
   \caption{Top, examples of hair swatches images acquired using our imaging device. Bottom, examples of synthetic images obtained with our hair path-tracing renderer.} 
\label{fig:real_vs_synth}
\end{figure}

Our contributions can be summarized as follows:

\begin{itemize}
\item We propose to combine controlled imaging and deep inverse graphics models to provide material capture solutions for non-differentiable renderers. 
\item We apply this method to create a hair sample digitization pipeline that can capture the appearance of a hair sample and synthesize images of hair with a similar appearance using a path-tracing renderer.
\item We validate our hair digitization model on synthetic and real-world data. 
\end{itemize}

\section{Method}

\subsection{Hair Strand Imagery}

The imaging setup used for capturing hair swatch images is illustrated in Figure \ref{fig:image_device}. It is composed of fixed camera and illumination to obtain controlled acquisition conditions. 
The hair swatch is stretched out on a flat-surface holder to fix the distance to the camera. A different geometry, such as a curved surface for holding strands, could be used to capture specular and secondary hair reflections.
Example images of hair acquired with this system are visible in Figure~\ref{fig:real_vs_synth}.

Compared to other hair capture systems \cite{paris2004capture,hu2014robust}, our acquisition method focuses on hair color and does not capture the geometry of a complete hairstyle. 
This choice was made to obtain a more scalable hair digitization approach, that can be done at a large scale using small synthetic/real hair swatches, which is more convenient for hair dying development.
Furthermore, the hair geometry can still be edited in the rendering environment, as illustrated in Figure \ref{fig:header_figure}.

\subsection{Hair Path-Tracing Renderer}
\label{sect:hair_renderer}


The hair renderer used in this approach is based on the scattering model described in~\cite{Yan15, Pharr201601TI}. Since the objective of this work is hair appearance, a physically-based renderer that accurately describes the lobes of light bounces on the hair cylinder is essential. The path tracing implementation of our renderer was performed in the parallelized framework \textit{Nvidia OptiX} \cite{Optix} to accelerate the computations. 

\begin{figure}[!]
\begin{center}
   \includegraphics[width=1.\linewidth]{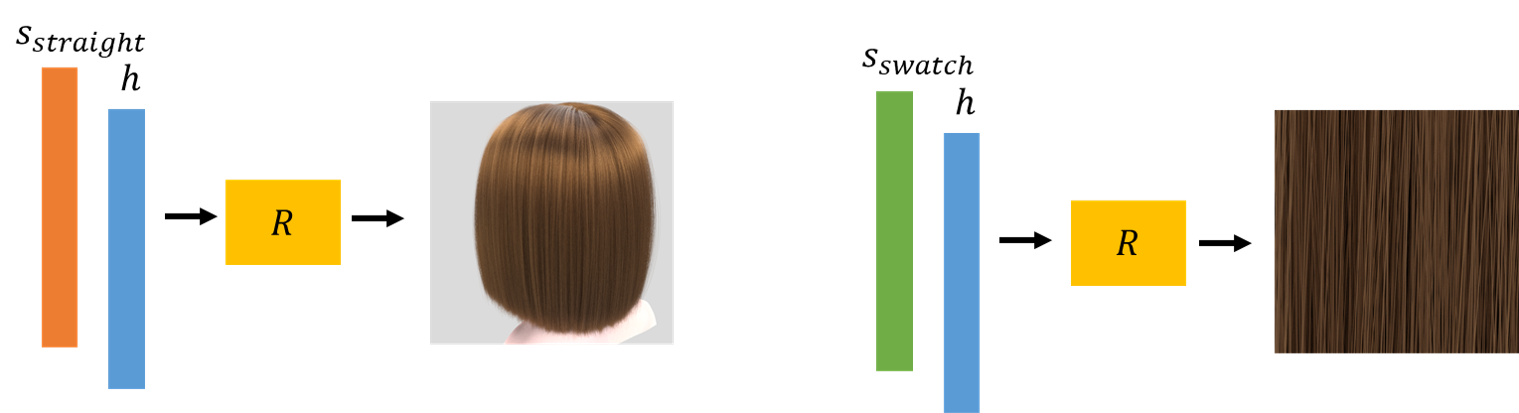}
\end{center}
   \caption{Our renderer takes as input hair parameters $h$ that determine the hair color and scene parameters $s$ which control camera position and hair shape.}
\label{fig:hair_scenes}
\end{figure}

As illustrated in Figure \ref{fig:hair_scenes}, we divide the rendering parameters into two categories: the hair parameters $h$ that determine the hair color and the scene parameters $s$ which control the other scene parameters such as camera position and hair shape. 
For a given set of hair parameters, the scene parameters can be dynamically controlled to produce images at different scales, and different hairstyles, as seen in Figure~\ref{tab:hair_param_desc}. The interaction with the renderer for controlling the hair color in the scene is performed by tuning the parameters of the concentration/ratio of the natural hair melanin of individual hair fibers~\cite{dEon11} as well as an additional color absorption parameter simulating an artificial dye colorant.

\begin{table}[h!]
\caption{Table 1: Description of the parameters controlling the hair color in our renderer.}
\begin{center}
\begin{tabular}{cc}
    \hline
    parameter & range \\
    \hline
    dye R,G,B & [0, 255] \\
    dye concentration & [0, 1] \\
    melanin concentration & [0, 1]\\
    melanin ratio (eumelanin/pheomelanin) & [0, 1] \\
    \hline
\end{tabular}
\end{center}
\label{tab:hair_param_desc}
\end{table}

\subsection{Hair Inverse Graphics Encoder}

\paragraph{Self-supervised training}

\begin{figure}[!]
\begin{center}
   \includegraphics[width=.8\linewidth]{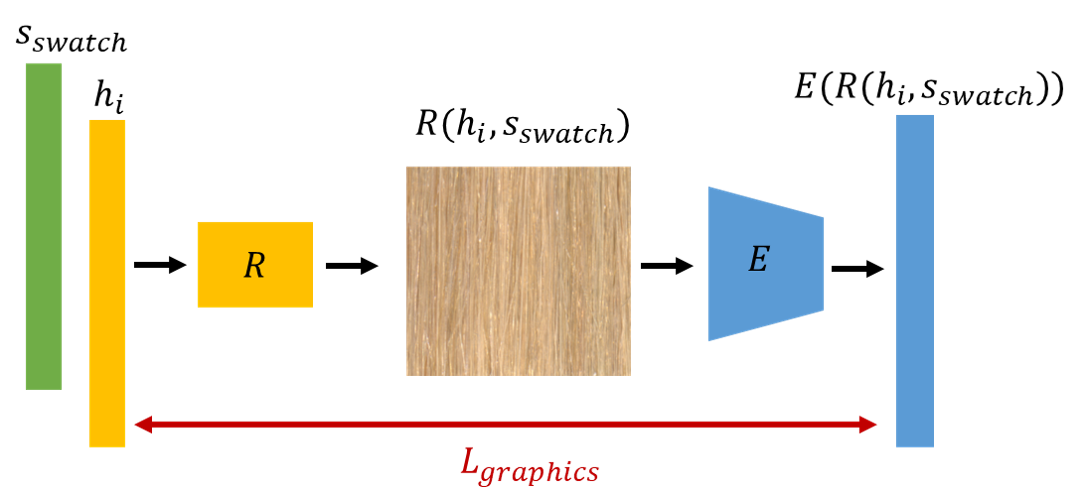}
\end{center}
   \caption{The training procedure of our hair inverse graphics model. Hair color parameters $h_i$ are randomly sampled and passed to the renderer $R$ to produce a synthetic swatch image using adapted scene parameters $s^{swatch}$. The rendered image is passed to the inverse graphics encoder $E$ that learns to estimate the initial hair parameters $h_i$ using $L_{graphics}$, a loss function defined in the graphics parameter space.}
\label{fig:model_training}
\end{figure}

Building upon the deep inverse graphics approach from
\cite{Kips_2021_CVPR}, we propose a deep inverse graphics model using self-supervised learning and synthetic images. 
In particular, this method allows learning an inverse graphics encoder without the need for a differentiable renderer, replacing the loss function defined in the image space by a loss function defined in the space of rendering parameters.
For each synthetic image $i$ we sample a random vector of hair parameters $h_i$, using the parameters described in Table~\ref{tab:hair_param_desc}. 
To obtain training with a large diversity of hair colors, we sample $n$ vectors of hair parameters using a uniform distribution for each parameter.
This sometimes leads to unrealistic hair colors, as seen in Figure~\ref{fig:real_vs_synth}, but ensures that our model generalizes well to rare hair colors such as blue or pink.
To introduce hair fiber localization variations among the synthetic images, we randomly sample for each synthetic image the camera position parameters, that are defined with spherical coordinates.
We denote this random scene parameters for synthesizing the swatch image~$i$ as $s^{swatch}_i$. 
These rendering parameters are then passed to the ray tracing renderer to produce the synthetic image $R(h_i, s^{swatch}_i)$. 
Finally, the synthetic image is given as input to an encoder network $E$, which is trained to estimate the initial hair parameters $h_i$. 
This training procedure is illustrated in Figure~\ref{fig:model_training}.
In total, the deep inverse graphics encoder $E$, parametrized by its weights $\Theta$, is trained to minimize the following loss function using gradient descent:
$$ \min_{\Theta} L_{graphics} = \min_{\Theta} \frac{1}{n}  \sum_{i=1}^n \left\lVert h_i - E(R(h_i, s^{swatch}_i)) \right\rVert ^2$$.

\begin{figure}[!]
\begin{center}
   \includegraphics[width=1.0\linewidth]{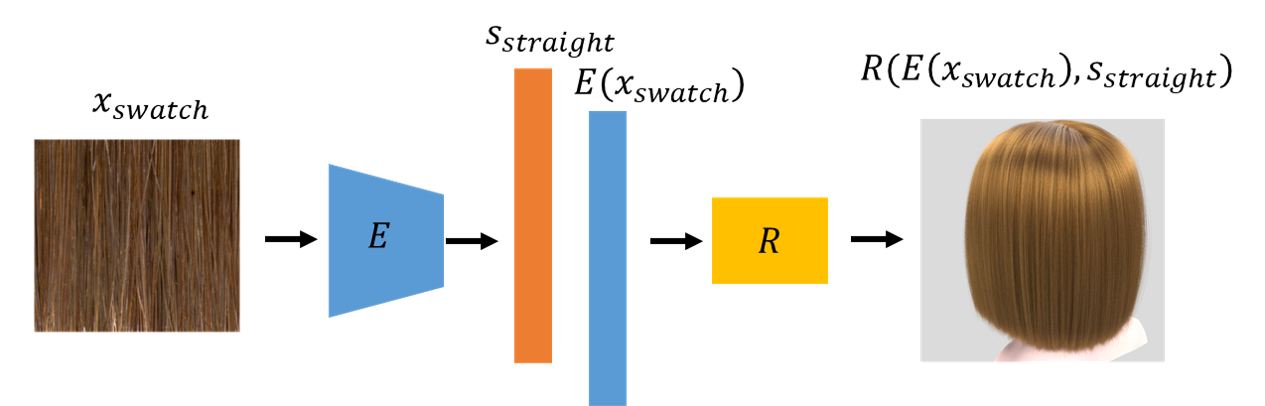}
\end{center}
   \caption{Our hair digitization pipeline. An image $x_{swatch}$ of the hair sample is acquired and passed to the inverse graphics encoder $E$ to estimate hair color parameters. 
   Synthetic images of hair with the same appearance as the hair sample can be rendered using the estimated hair parameters and the chosen scene parameters $s$. In this example, $s_{straight}$ allows synthesizing portrait scale images of straight hair.}
\label{fig:inference_pipeline}
\end{figure}

\begin{figure*}[t!]
\begin{center}
   \includegraphics[width=0.56\linewidth]{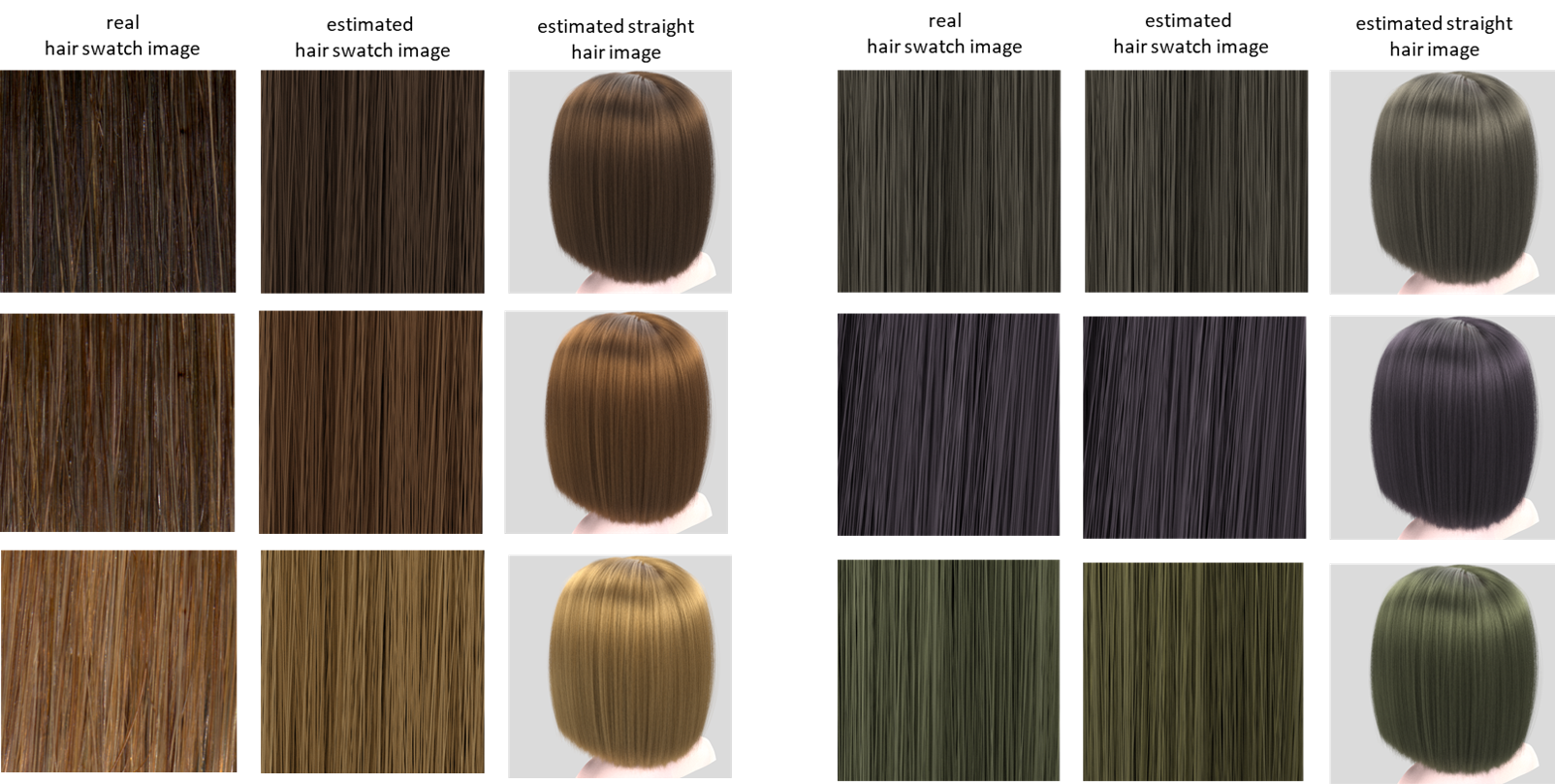}
\end{center}
   \caption{Qualitative results on real and synthetic hair swatch images. Our inverse graphics model accurately captures the hair appearance from hair swatch images, over a large variety of hair colors.
   }
\label{fig:quali_real}
\end{figure*}

\paragraph{Hair digitization}

At inference time, the inverse graphics encoder is used to estimate hair color using real hair swatches. 
Our imaging system is used to obtain an image of a hair sample, that we denote $x_{swatch}$. 
This image is sent to the graphics encoder which estimates the associated hair rendering parameters $E(x_{swatch})$.
Finally, these hair parameters can be used with the path-tracing engine to render synthetic images with various scene parameters $s$, such as different hairstyles, lighting conditions, or camera positions. This hair digitization pipeline is illustrated in Figure~\ref{fig:inference_pipeline}.

\section{Experiments and Results}

\subsection{Implementation}
To train our graphics encoder we use the same fully convolutional neural network architecture as in~\cite{Kips_2021_CVPR}. We use a training set of $n=5000$ synthetic images generated using the renderer described 
previously.
Our model is trained over 400 epochs using the Adam optimizer, 
a fixed learning rate of $10^{-6}$ and a batch size of 32.
Since existing methods cannot capture the appearance of a hair sample but only function on a complete hair, we cannot compare our approach to existing baselines.

\subsection{Qualitative Evaluation}

We performed qualitative experiments on both synthetic and real data.
First, we captured images of real hair swatches using our imaging system and with our inverse graphics approach we synthesized images with various scene settings, as seen in Figure~\ref{fig:header_figure}.
For assessing performance on more challenging hair colors, we synthesized a set of test swatch images and repeated the experiment.
For both real and synthetic images, it can be observed that our approach allows us to accurately capture the hair appearance over a range of various hair colors. 
The fine color variations between several shades of brown hair are still visible in the rendered images, which tends to show that our model is accurate enough to be used in practice.

However, it can be observed that our model is not able to reproduce the natural hair color variation that can exist within individual hair fibers. 
This limitation could be overcome by introducing heterogeneity in hair parameters, both in the renderer and inverse graphics model, such as a standard deviation for each color parameter. 

\subsection{Quantitative Evaluation}

In order to quantitatively assess the performance of our approach, we also performed synthetic experiments.
We synthesized a set of 300 original hair swatch images using random hair parameters drawn according to a uniform distribution. For each synthetic image, we estimated the hair parameters using our inverse graphics encoder, and rendered the corresponding image using the same scene parameters to obtain images with aligned hair fibers. 
Finally, we computed various image reconstruction metrics between the original and the reconstructed strand images. The results of this experiment are reported in Table~\ref{tab:image_rec_metric}. The low errors on all image reconstruction metrics tend to confirm the qualitative evaluation results. 

\begin{table}[!]
\caption{Table 2 : Hair image reconstruction performance on synthetic data}
\begin{center}       
\begin{tabular}{c c}
\hline
    evaluation measure &  value (mean $\pm$ std )\\
    \hline
    L1 & 7.08 $\pm$ 11.79 \\
    \hline
    MSSIM \cite{wang2003multiscale} & 0.20 $\pm$ 0.04 \\
    \hline
    LPIPS \cite{zhang2018unreasonable} & 0.12 $\pm$ 0.19 \\
    \hline
\end{tabular}
\end{center}
\label{tab:image_rec_metric}
\end{table}





\section{Conclusion}
In this paper, we introduced a novel hair 
method for automatic digitization of hair color appearance between real samples and synthetic images. Both qualitative and quantitative results imply satisfactory results for real applications.

Future work might seek to include color heterogeneity of hair fibers and varying specular properties (hair shine) adapting the imaging and the rendering system.



\printbibliography




\end{document}